\documentclass[aps,prb,preprint,showpacs,floatfix]{revtex4-1}
\usepackage{amsmath}
\usepackage{amssymb}
\usepackage{graphicx}
\usepackage{xcolor}

\newcommand{\figref}[1]{Fig.~\ref{#1}}

\newcommand{\secref}[1]{Section~\ref{#1}}

\newcommand{\bra}[1]{\ensuremath{\langle{#1}|}}
\newcommand{\ket}[1]{\ensuremath{|{#1}\rangle}}
\newcommand{\bracket}[2]{\ensuremath{\langle{#1}|{#2}\rangle}}

\newcommand{\proj}[1]{\ensuremath{\ket{#1} \bra{#1}}}

\begin{document}

\title{Optimizing phase estimation algorithms for diamond spin magnetometry}
\author{N. M. Nusran}
\author{M. V. Gurudev Dutt}
\email[Correspondence to:]{gdutt@pitt.edu}
\affiliation{Department of Physics and Astronomy, University of Pittsburgh, Pittsburgh, Pennsylvania 15260, USA}	

\date{\today}

\begin{abstract}
We present a detailed theoretical and numerical study discussing the application and optimization of phase estimation algorithms (PEAs) to diamond spin magnetometry. We compare standard Ramsey magnetometry, the non-adaptive PEA (NAPEA) and   quantum PEA (QPEA) incorporating error-checking. Our  results show that the NAPEA requires lower measurement fidelity, has better dynamic range, and greater consistency in sensitivity. We elucidate the importance of dynamic range to Ramsey magnetic imaging with diamond spins, and introduce the application of PEAs to time-dependent magnetometry.  
\end{abstract}
\maketitle


\section{Introduction}
\label{sec:intro}

Quantum sensors in robust solid-state systems offer the possibility of combining the advantages of precision quantum metrology with nanotechnology. Quantum electrometers and magnetometers have been realized with superconducting qubits~\cite{Bylander11}, quantum dots~\cite{Vamivakas11}, and spins in diamond~\cite{Maze08,Taylor08,Balasub08,Dolde11}. These sensors could be used for fundamental studies of materials, spintronics and quantum computing as well as applications in medical and biological technologies. In particular, the electronic spin of the nitrogen-vacancy (NV) color center in diamond has become a prominent quantum sensor due to optical transitions that allow for preparation and measurement of the spin state, stable fluorescence even in small nanodiamonds\cite{Bradac10}, long spin lifetimes\cite{Balasub09}, biological compatibility\cite{CCFu07,McGuinness11} as well as available quantum memory that can be encoded in proximal nuclear spins\cite{Jelezko04b,Dutt07}.  

The essential idea of quantum probes is to detect a frequency shift $\delta \nu$ in the probe resonance caused by the external perturbation to be measured. The standard method to do this with maximum sensitivity is the Ramsey interferometry scheme, which measures the relative phase $\phi = \delta \nu \times t$ accumulated by the prepared superposition of two qubit states $(\ket{0} + \ket{1})/ \sqrt{2}$. These states will evolve to $(\ket{0} + e^{-i \phi} \ket{1})/ \sqrt{2}$, and subsequent measurement along one of the two states will yield a probability distribution $P(\phi) \sim \cos (\phi)$, allowing the frequency shift $\delta \nu$ to be measured. For NV centers the detuning $\delta \nu = \gamma_e B $ where $\gamma_e \approx 2 \pi (28)$~GHz/T is the NV gyromagnetic ratio and $B$ is the field to be measured. 

The phase (or field) sensitivity is obtained by assuming that the phase has been well localized between the values $(\phi-\pi/2, \phi+\pi/2)$, where $\phi$ is the actual quantum phase value, and in practice to much better than this by making a linear approximation to the sinusoidal distribution. Thus, prior knowledge of the ``working point'' of the quantum sensor is key to obtaining the high sensitivity that makes the sensors attractive. When the actual phase $\phi$ is allowed to take the full range of values, then the quantum phase ambiguity (i.e. the multi-valued nature of the inverse cosine function) results in much larger phase variance than predicted by the standard methods. To overcome the quantum phase ambiguity, we require an estimator $\phi_{est}$ that can achieve high precision (small phase variance) over the entire phase interval $(-\pi, \pi)$ without any prior information. In terms of field sensing, this translates to a high dynamic range for magnetometry, i.e. to increase the ratio of maximum field strength ($B_{max}$) to the minimum measurable field ($\delta B_{min}$) per unit of averaging time. This would be a typical situation in most applications of nanoscale magnetometry and imaging, where unknown samples are being probed. In fact, as we show in \secref{sec:dynrng}, as the sensitivity increases, it will be increasingly difficult to image systems where there is more than one type of spin present, and errors in the NV position will result in significant errors in the mapping. Further, since Ramsey imaging results in only one contour of the field being mapped out in a given scan, the acquisition time is greatly reduced, and thus several images must be made to accurately reconstruct the position of the target spins.

Recently, phase estimation algorithms (PEA) were implemented experimentally with both the electronic~\cite{Nusran12} and nuclear spin qubits~\cite{Waldherr12} in diamond to address and resolve the dynamic range problem.  While we note that the theory for the nuclear spin qubit has been presented in Ref.~\cite{Said11}, our work supplements this by applying the theory to the electronic spin qubit which is more commonly used for magnetometry. Some of the questions that we address in this work, and have not been studied earlier, include: (i) the importance of control phases in the PEA, (ii) the dependence of sensitivity on the control phases, (iii) the dependence of sensitivity on the dynamic range, and (iv) the impact of measurement fidelity on the PEAs. We have also studied the application of the PEA to magnetometry of time-dependant fields, and demonstrate the usefulness in measuring both amplitude and phase of an oscillating (AC) magnetic field. 

In \secref{sec:qmag}, we present a brief introduction and overview of the NV center, Ramsey interferometry, and the importance of dynamic range in magnetic imaging. \secref{sec:pea} introduces the two types of PEAs that we have compared in this work. \secref{sec:results} shows some of the important results we obtain through the simulations. This includes a discussion on the importance of control phases, weighting scheme, required measurement fidelity and the possibility of implementing PEA for phase-lockable AC magnetic field detection. Finally, \secref{sec:conclusion} summarizes the conclusions.

\section{Diamond quantum magnetometry}
\label{sec:qmag}

The NV ground state consists of a $S = 1$ spin triplet, with a natural quantization axis provided by the defect symmetry axis between the substitutional nitrogen and adjacent vacancy that constitute the color center. In the absence of a magnetic field, the ground state $m_s = 0$ spin sublevel is split by $D = 2 \pi (2.87)$~GHz from the $m_s = \pm 1$ levels. By applying a small magnetic field, the magnetic dipole moment of the NV causes the states $\ket{\pm 1}$ to split. The optically excited state of the NV defect also has the triplet $S  = 1$ configuration, oriented along the same quantization axis and with similar magnetic moment, but its room temperature zero-field splitting is only $2 \pi (1.42)$~GHz\cite{Fuchs08,Neumann09}. 

In the experiments of Ref.\cite{Nusran12}, a static magnetic field of 40 mT is applied along the NV axis by a permanent magnet. Level anti-crossing (LAC) occurs between the $\ket{m_S = -1}$ and $\ket{m_S = 0}$ sublevels in the excited state, which results in dynamic nuclear-spin polarization (DNP) of $^{14}N$ nuclear spin (I=1) associated with most NV experiments ~\cite{Jacques09}. Microwave is delivered via a 20 micron diameter copper wire placed on the diamond sample, which is soldered into an impedance matched strip-line. The resonant microwave radiation for the $\ket{m_S = 0} \leftrightarrow \ket{m_S = -1}$ transition leads to coherent manipulation of the spin. Due to the wide separation of the electronic spin levels and the DNP mechanism, we can treat the system as a pseudo-spin $\sigma = 1/2$ qubit, and write the Hamiltonian in the rotating frame, with the rotating wave approximation as $H = \hbar \delta \nu \sigma_z + \hbar \Omega (\sigma_x \cos \Phi + \sigma_y \sin \Phi)$ where $\Omega$ is the Rabi frequency, and $\Phi$ is the control phase. Our theoretical simulations below assume typical numbers from the experiments of Ref.~\cite{Nusran12}, but we shall discuss the consequences of improved experimental efficiency where appropriate. For clarity, we shall use the notations $\ket{0} \equiv  \ket{m_S = 0}$ and $\ket{1} \equiv \ket{m_S = -1}$ to describe the qubit basis states from here onwards.

\subsection{Standard measurement limited sensitivity}
\label{sec:sql}
 
\begin{figure}[bht]
	\scalebox{1}[1]{\includegraphics[width=5in]{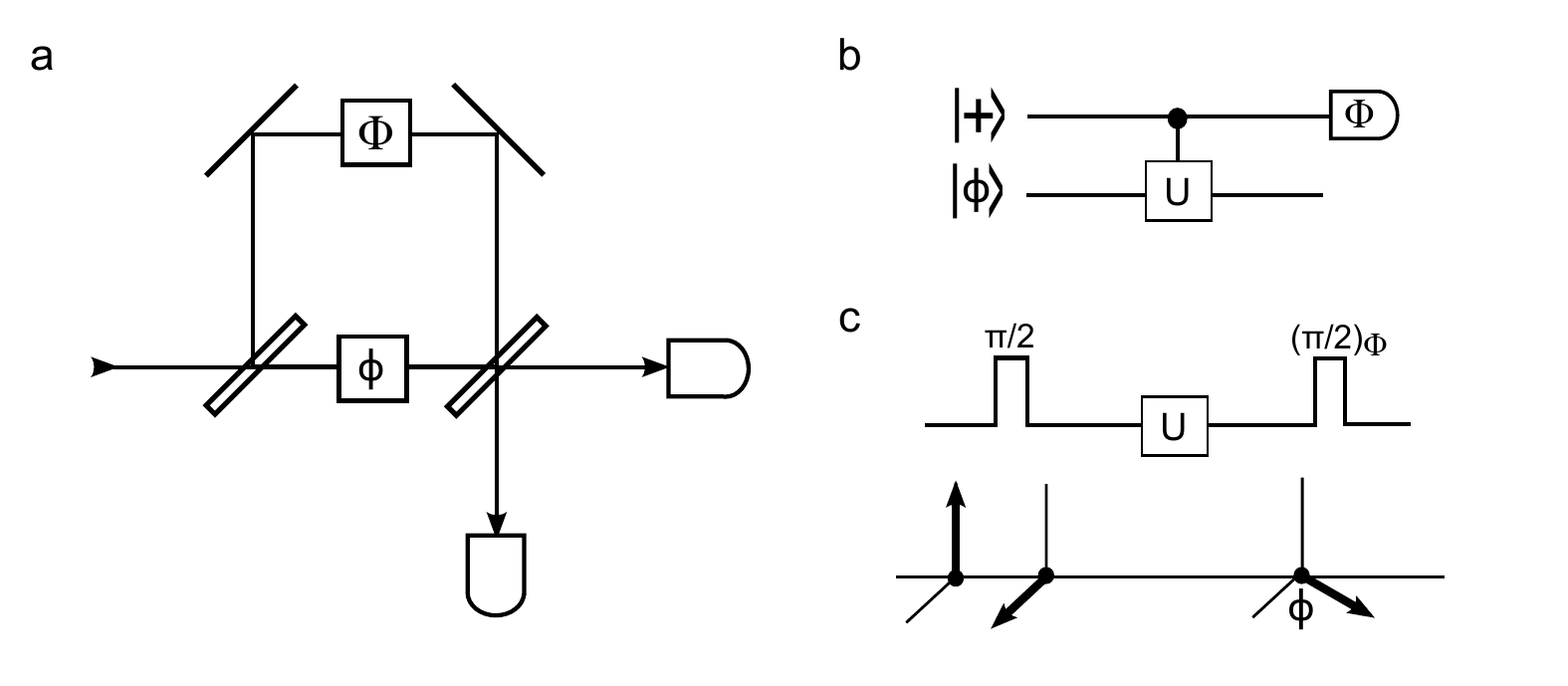}}
	\caption{\small{\textbf{Quantum metrology:} \textbf{a,} Mach-Zehnder interferometer senses the relative phase shift between two beams paths. \textbf{b,} Quantum circuit representation of the process and \textbf{c,} The analogous realization via Ramsey interferometry. The first $\pi/2$-pulse brings the spin vector from $\ket{0}$ to $(\ket{0}+\ket{1})/\sqrt{2}$. The free introduces a phase $\phi$ on $\ket{1}$ relative to $\ket{0}$. Measurement after the second $\pi/2$-pulse senses this phase with respect to the control phase $\Phi$.}}
	\label{fig:mzi}
\end{figure}

The standard model for phase measurements in quantum metrology is depicted in \figref{fig:mzi}. The equivalence of Mach-Zehnder interferometry (MZI) depicted in \figref{fig:mzi}(a), the quantum network model (QNM) in \figref{fig:mzi}(b)\cite{Cleve98}, and Ramsey interferometry (RI) \figref{fig:mzi}(c) for phase estimation  allows us to treat all three problems in a unified framework. Thus, it was pointed out by Yurke et al \cite{Yurke86} that the states of photons injected into the MZI can be rewritten through application of Schwinger double-ladder operators to represent spin states.  They showed that the number phase uncertainty relation for photons could be derived from the angular momentum commutation relations. Similarly, in the quantum network model \cite{Cleve98}, auxiliary qubits (or classical fields) are prepared in an eigenstate of the operator $U$ such that $U\ket{\phi} = e^{-i \phi} \ket{\phi}$. The controlled-U operations on the system results in the following sequence of transformations on the qubits,
\begin{equation}
\ket{+}\ket{\phi} \xrightarrow{c-U}  \frac{\ket{0} + e^{-i \phi} \ket{1}}{\sqrt{2}} \ket{\phi} \xrightarrow{H} \left(\cos(\frac{\phi}{2})\ket{0} + i \sin(\frac{\phi}{2}) \ket{1} \right) e^{-i \phi/2} \ket{\phi}
\end{equation}
The state of the auxiliary register, being an eigenstate of $U$, is not altered along the network, but the eigenvalue $e^{-i \phi}$ is ``kicked-out'' in front of the $\ket{1}$ component of the control qubit. This model has allowed for application of ideas from quantum information to understand the limits of quantum sensing (see Refs.\cite{Huelga97, Giovannetti04,Giovannetti06}).

Quantum metrology shows that the key resource for phase estimation is the number of interactions of $n$ spins with the field prior to measurement ($n$ measurement passes). Classical and quantum strategies differ in the preparation of uncorrelated or entangled initial states, respectively.  Parallel and serial strategies differ in whether after the initial preparation, all spins are treated identically in terms of evolution and measurement. Thus, a serial strategy can trade-off running time with number of spins to achieve the same field uncertainty. In either case, the limiting resources can be expressed in one variable: the total interaction time $T = n t$. While quantum strategies can in principle achieve the Heisenberg phase uncertainty $ \langle (\Delta \phi)^2 \rangle  \propto 1/n^2 \propto 1/ T^2$, classical strategies (whether parallel or serial), however, can at best scale with the phase uncertainty  $ \langle (\Delta \phi)^2 \rangle  \propto 1/n \propto 1/ T$.  This limit, known as the standard measurement sensitivity (SMS) arises from the combination of two causes: the probabilistic and discrete nature of quantum spin measurements, and the well-known central limit theorem for independent measurements~\cite{Giovannetti04,Giovannetti06}. 

However, in obtaining the phase from the number of spins found to be pointing up or down after a measurement, there is an ambiguity. Because of the sinusoidal dependence of the phase accumulated, the above expression for the SMS assumes the phase has already been localized to an interval $(-\pi/2, \pi/2)$ around the true value. But for unknown fields, the entangled states typically accumulate phase $\sim n$ times faster than a similar un-entangled state. Thus, the working point must be known much more precisely for such strategies to be successful, which may defeat the original purpose of accurate field estimation. Thus, such quantum entangled strategies are better suited for situations where there are only likely to be small changes from a previously well-known field. 

Consider the case of the classical strategy: we can write the interaction time $T = n T_2$, where $T_2$ is the decoherence time, and since $\delta \phi \sim 1/ \sqrt{n}$ we obtain the field uncertainty $ \delta B \sim \frac{1}{\gamma_e \sqrt{T_2 T}}$. This also implies that the field must be known to lie in the range $\vert B \vert \leq B_{max} = \frac{\pi}{2 \gamma_e T_2}$. Putting these together, we have that the dynamic range $DR = \frac{B_{max}}{\delta B} \sim \frac{\pi}{2} \sqrt{\frac{T}{T_2}}$. Thus the dynamic range will decrease as the coherence time increase.

The above expressions for the SMS do not take into account the effects of decoherence, measurement imperfections and other types of noise in experiments. We use the density matrix approach to describe the state of the quantum system and take into account these effects. Any unitary interaction on a single spin is essentially a rotation in the Bloch sphere. If we assume that the Rabi frequency $\Omega \gg \delta \nu$, we can assume that the rotations are instantaneous, and neglect the effect of the free evolution during the time of the pulses. In numerical simulations, we could also include the effect of finite $\Omega$ and $\delta \nu$ easily. For instance, a simple Ramsey experiment could be simulated as follows: an initial density matrix $\rho_0 = \ket{0}\bra{0}$ is first brought to $\rho_1 =  R_y(\pi/2) \rho_0 R_y(\pi/2)^{\dagger}$, where $R_n(\theta) = \exp{(-i (\vec{\sigma} . \vec{n}) \theta /2)}$ is the rotation operator along the $\hat{n}$ direction. This is equivalent to the action of a  $(\pi/2)_y$ pulse in the experiment. Letting the system to evolve freely under the external magnetic field leads to the state:  $\rho_t =  U(t) \rho_1 U(t)^{\dagger}$, where $U(t)=\exp{(- i \delta \nu \sigma_z t)}$ is the time evolution operator. The application of the final $(\pi/2)_{\Phi}$ pulse is achieved using a z-rotation followed by $R_y(\pi/2)$: 
\[ \rho_f = R_y(\pi/2) R_z(\Phi) \rho_t R_z(\Phi)^{\dagger} R_y(\pi/2)^{\dagger} \]
The effect of decoherence is introduced by multiplying the off-diagonal elements with the decay factor $D(t,T_2^*)=\exp{(-(t/T_2^*)^2)}$, where $T_2^*$ is the dephasing time set in our simulations. The probability for the measurement of the state in the Ramsey experiment is then given by\cite{Said11}, 
\begin{equation}
P(u_m | \phi) = \frac{1 +(-1)^{u_m}  D(t, T_2^*) \cos (\phi - \Phi)}{2}
\label{eq:probumphik}
\end{equation}
where measurement bit $u_m = 1 (0)$ is applied to to state $\ket{0}(\ket{1})$. The bit $u_m$ is determined by comparing the measured signal level with a pre-defined threshold value. Further, feedback rotations $\Phi$ are simply achieved by controlling the phase of the second $\pi/2$ microwave pulse.
Repeating the experiment $n$ times, we obtain the fraction of spins $n_0 (n_{1})$ that actually point up or down, thereby inferring the probability, e.g. $P(u_m = 1 | \phi) = \frac{n_0}{n}$. The last step is to take the inverse of this equation and obtain $\phi$. Unfortunately, as pointed out earlier, the inverse cosine is multi-valued, and thus we have the quantum phase ambiguity which requires us to have prior knowledge about the phase and the working point for the Ramsey experiments.

The SMS limit can be calculated for our Ramsey experiments with NV centers, using the definition $(\delta \phi)^2 = \frac{\langle (\delta S)^2 \rangle}{\lvert d S / d \phi \rvert^2}$ i.e., by assuming that signal to noise ratio $SNR = 1$. Because the phase error $\delta \phi = \gamma_e \delta B t$, we can also calculate the sensitivity $\eta = \delta B \sqrt{T}$. Here, $S =  \langle Tr (M \rho) \rangle$ represents the signal from Ramsey experiments, and the variance of the signal, $\langle (\delta S)^2 \rangle =  Tr (M^2 \rho) - (Tr( M \rho))^2 $. The optical measurement operator $M = a \ket{0}\bra{0} + b \ket{1}\bra{1}$, where $a, b$ are Poisson random variables with means  $\kappa \alpha_0 $ ($ \kappa \alpha_1$) that represent our experimental counts when the qubit is in the $\ket{0}$ ($\ket{1}$) state respectively. 
Here, $\alpha_0$ and $\alpha_1$ represent the photon counts per optical measurement shot and $\kappa$ is the number of times the measurement is repeated till the qubit state can be distinguished with sufficiently high fidelity, $f_d$. For instance, standard quantum discrimination protocols imply that $f_d >0.66$ is sufficient to distinguish unknown pure states from a random guess~\cite{Massar95}. The value of $\kappa$ can be tuned in the simulations and experiments, but after fixing $\kappa$ for a given experiment, $N$ is then simply the statistical repetitions needed to find the system phase $\phi$, thus the number of resources $ n = N \cdot \kappa$. In the limit of single-shot readout $\kappa = 1$ on the electronic spin state, it is clear that quantum projection noise limits for $n$ and $N$ are equivalent, and otherwise they are proportional by a scale factor that depends on experimental efficiency.  

We can explicitly calculate the sensitivity (with $SNR = 1$) for Ramsey measurements for general working points. From definitions, it can be shown analytically that 
\begin{equation}
\eta^2 = 
\frac{\kappa_{th} }{\gamma_e^2  t D(t, T_2^*)^2 \sin^2(\phi - \Phi)} 
\label{eq:etaSMS}
\end{equation}
with,
\begin{equation}
\kappa_{th} =   1 + 2  \frac{\alpha_0 + \alpha_1}{(\alpha_0 - \alpha_1)^2} +  \frac{2 D(t, T_2^*) \cos(\phi - \Phi)}{\alpha_0 - \alpha_1} - D(t, T_2^*)^2 \cos^2(\phi - \Phi) 
\end{equation}
Similar results have also been derived by Refs.~\cite{Taylor08,Meriles10}. This expression reduces to the ideal SMS $\eta_{ideal}^{SMS} = \frac{e^{0.5}}{\gamma \sqrt{T_2^*}}$ in the limit of perfect experimental efficiency ($\kappa_{th} =1$), and assuming that $\phi = 0, \Phi = \pi/2$. The importance of the working point can clearly be seen in this derivation since small changes in $\phi$ from the working point result in quadratic increase of $\eta$. The factor $\kappa $ may be thought of as a loss mechanism, i.e. when we repeat the experiment $N$ times,  we only gain information from $1/\kappa$ of the runs during measurement and hence we must repeat the experiment $\kappa$ times to achieve the same sensitivity. For ideal (single-shot) measurements, which could potentially be realized through resonant excitation and increased collection efficiency\cite{Babinec10}, the SMS is given by taking $\alpha_1 / \alpha_0 \rightarrow 0$ and $\alpha_0 \gg 1$, resulting in $\kappa = 1$.

\begin{figure}[bht]
	\scalebox{1}[1]{\includegraphics[width=5in]{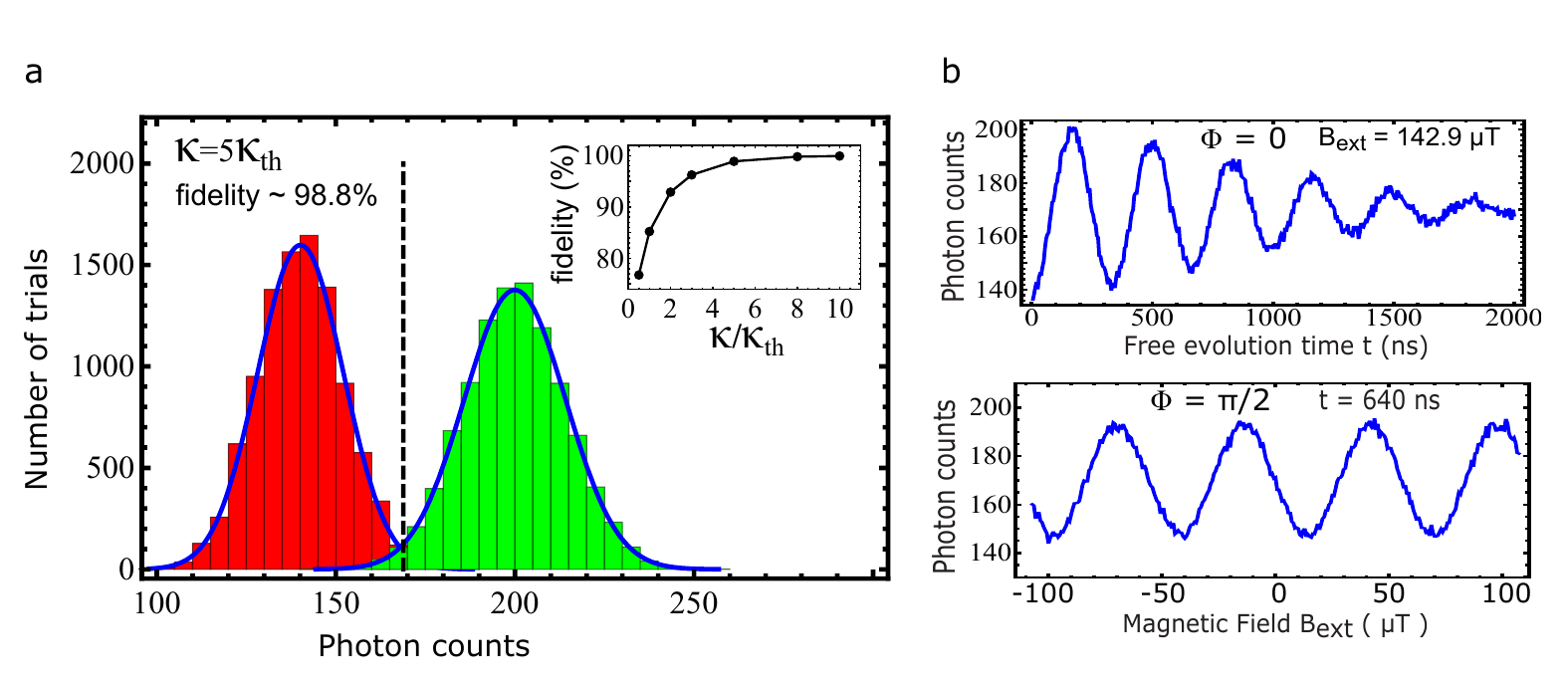}}
	\caption{\small{\textbf{Ramsey Simulations:} \textbf{a,} Measurement fidelity: The histogram obtained with $N =10000$ measurement for both $\ket{0}$ and $\ket{1}$ states are shown. Lesser the overlap higher the fidelity in distinguishing the states. A sample number $\kappa =5\kappa_{th}$ leads to a fidelity $\approx 98.8\%$. (inset) Fidelity improves with $\kappa$.  \textbf{b,} Simulation of Ramsey Experiment: Ramsey signal after 50 averages when free time $t$ is varied while $B_{ext} = 142.9 \, \mu$T is fixed (top) and as $B_{ext}$ varied while $t = 640$~ns is fixed (bottom). Here, $\kappa =5\kappa_{th}$ was used.}}
	\label{fig:simmethod}
\end{figure}

To verify our simulation method needed for the phase estimation algorithms, we first carried out Monte-Carlo simulation procedure for Ramsey fringes where we have the analytical results derived above for comparison. \figref{fig:simmethod} demonstrates first the measurement fidelity for distinguishing  $\ket{0}$ and $\ket{-1}$ states for two different sample sizes. The measurement fidelity is defined as 
\[f_d = \frac{\bracket{0}{\rho^{0} | 0} + \bracket{1}{\rho^{1}|1}}{2} \]
where $\rho^{0(1)}$ corresponds to the state initially prepared in $\ket{0}$($\ket{1}$).
The average photon counts per optical measurement for the state $\ket{0}$ ($\ket{1}$) $\alpha_0 = 0.010$ ($\alpha_1 =0.007$) are set throughout our simulations to correspond with the experiments of Ref.\cite{Nusran12}. The experimental sequence was run with initialization of the spin into the $\ket{0}$ state, followed either immediately by fluorescence measurement or by a $\pi$ pulse and then measurement. The experimental threshold for the bit measurement $u_m$ is usually chosen as the average of the means of the two histograms. Our results, shown in \figref{fig:simmethod}(a), show that by tuning the number of samples $\kappa$, we can achieve very high measurement fidelity. 

\subsection{Impact of dynamic range on magnetic imaging}
\label{sec:dynrng}

\begin{figure}[hbt]
\centering 
\scalebox{1}[1]{\includegraphics[width=5in]{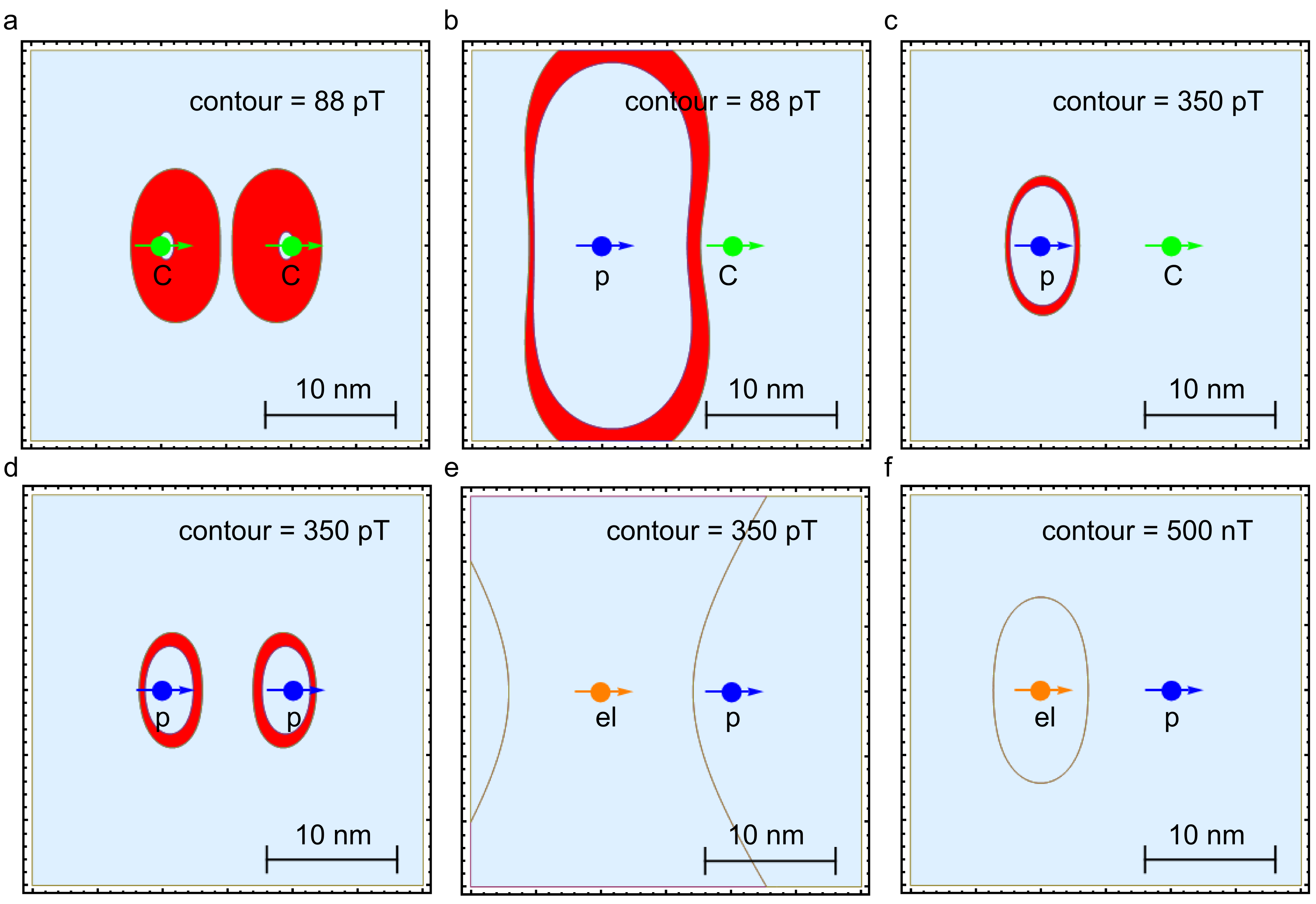}}
	\caption{\small{\textbf{Ramsey imaging:} \textbf{a-f,} Calculated resonant imaging contours by fixing the detuning of the Ramsey sequence at a value expected from the field of a single electron (el), proton (p) or carbon-13 (C) spin. The line-width of the sensor is set to be 30~pT which could be achieved with coherence time $T_2=100$~ms and averaging time $T=14$~s. The sample-probe height is 10~nm. The resonance condition will be met when the field from the target spins projected on the NV axis is within one linewidth $\delta B / \gamma_e$ of the contour value (Ramsey detuning), and the corresponding pixel in the image is shaded red to represent a dip in the fluorescence level. The contour images do not sufficiently reveal the existence of different spins. For instance, existence of carbon-13 is not revealed in \textbf{b} and \textbf{c} whereas the existence of proton is not revealed in \textbf{f}.}}
	\label{fig:imaging}
\end{figure}

In the standard approach of magnetic imaging, the contour height (Ramsey detuning $\delta \nu$) is set by estimating what the expected field would be at the NV spin, and calculating the corresponding detuning. The resonance condition will be met when the field from the target spins projected on the NV axis is within one linewidth $\delta B/\gamma_e$ of the Ramsey detuning, and the corresponding pixel in the image is shaded to represent a dip in the fluorescence level. Thus, only a single contour line of the magnetic field is revealed as a ``resonance fringe'' and a sensitivity limited by the intrinsic linewidth $1/T_2^*$ could be achieved. However, a quantitative map of the magnetic field is impossible in a single run due to the restriction of a single contour\cite{Haberle13}. In order to further illustrate the importance of dynamic range, we show in \figref{fig:imaging} the contours obtained for Ramsey imaging with a single NV center placed at a distance of 10~nm above different types of spins that are separated by $10$~nm. The contours are calculated by using the expressions for magnetic fields from point dipoles with magnetic dipole moments for the corresponding nuclear spin species. Our simulations do not take into account any measurement imperfections such as fringe visibility, and assume that the decoherence time of the NV spin can be made sufficiently long enough to detect the various species of spins, e.g. $T_2 = 100$~ms. From the figures, it is clear that when spin species of different types are present in the sample, the contours get greatly distorted and makes it difficult to reconstruct the position of the spin. 

Using the same procedures, we find that an error in the NV spin position inside the diamond lattice will significantly affect both resolution and image reconstruction. First, the resolution of the image is set by the gradient of the field from the target spin $\nabla B$ and the line width $\delta B$, giving rise to a resolution $\Delta x_{res} = \delta B / \nabla B$. Here $\nabla B = \frac{3\mu_0 m}{4 \pi r^4}$, where $m$ is the magnetic dipole moment of target spin, $\mu_0$ is the permeability in vacuum. From the prior expression for the line width, this becomes $\Delta x_{res} \sim\ \frac{\kappa}{\gamma_e \nabla B \sqrt{T_2 T}}$. For a line width $\delta B \approx 30$~pT and a target proton spin, we get $\Delta x_{res} \sim 0.75$~nm, which agrees well with the contour plots in \figref{fig:imaging}. Secondly, if the NV position has an error of $\delta r$, the working point will shift by $\nabla B \, \delta r$. When the shift is comparable to the field sensing limit of Ramsey measurements $B_{max}$ defined earlier,  $\frac{\pi}{2 \gamma_e T_2}$, we will lose the sensitivity needed to reconstruct the position. Putting in the numbers used for our simulations, we obtain that position reconstruction will not be possible if $\delta r \sim \frac{2 \pi^2 r^4}{3 \gamma_e \mu_0 m T_2} \approx 3.7$~nm. In practice the error in working point should be a fraction e.g. 30 -- 50\% of the dynamic range $B_{max}$ since the linewidth will be broadened otherwise. Under normal growth or implantation conditions for near-surface NV centers, the typical uncertainty in NV position is $\sim 1-10$~nm\cite{Mamin13, Staudacher13}. By increasing the dynamic range of the imaging technique, we can relax the requirement for knowing the NV position more accurately.

\section{Phase Estimation Methods}
\label{sec:pea}

\begin{figure}[hbt]
	\scalebox{1}[1]{\includegraphics[width=5in]{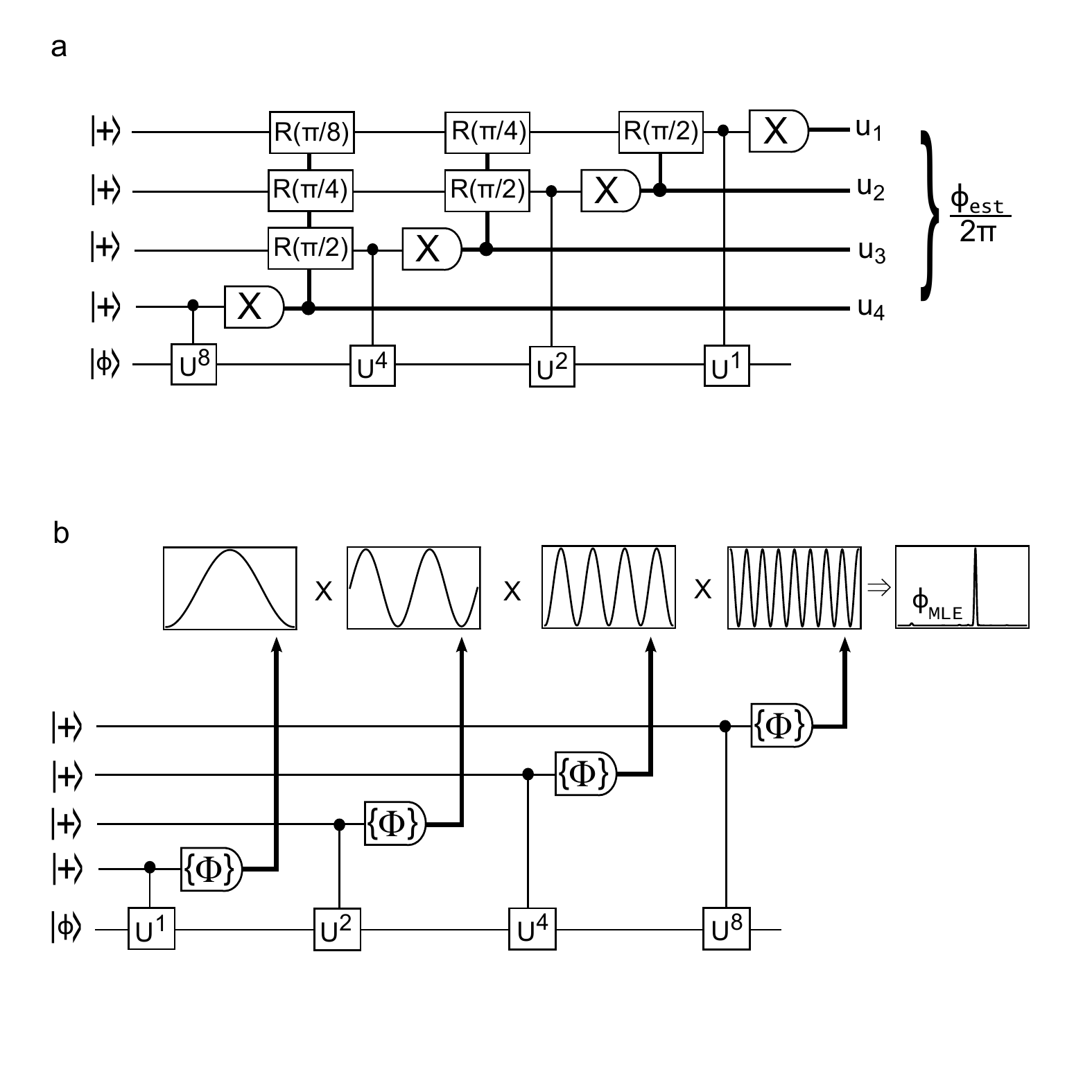}}
	\caption{\small{\textbf{Quantum circuit diagrams:} \textbf{a,} QPEA \textbf{b,} NAPEA for the case K=4. The dark lines represent the classical bits. The measured bit string in QPEA it self gives an estimate  $\phi_{est}$ for the unknown phase with $u_1$ being the least significant bit. The control rotations $R(\Phi)$ in QPEA depends on preceding measurement results. In contrast, measurements in NAPEA are performed with a pre-defined set of control phases $\{\Phi\}$ and Bayes' method is used to obtain $\phi_{MLE}$.}}
	\label{fig:algo}
\end{figure}

The quantum network model for quantum metrology allows us to apply ideas from quantum information to resolve the problem of dynamic range. To see how this works, let us consider the quantum phase estimation algorithm (QPEA) that utilizes the inverse quantum Fourier transform of Shor's algorithm. The QPEA (\figref{fig:algo}(a)) requires $K$ number of unitaries $(U^p, p=2^0,2^1,...,2^{K-1})$ to be applied in order to obtain an estimation 
\[\phi_{est} = 2 \pi (\frac{u_1}{2^0} + \frac{u_2}{2^1} \ldots +\frac{u_K}{2^{K-1}})\equiv 2 \pi (0. u_1 u_2 \ldots u_K)_2 \]
for the classical phase parameter $\phi$ , with $K$ bits of precision. When the phase is expressible exactly in binary notation (i.e. a fraction of a power of two), the QPEA gives an exact result for the phase estimator $\phi_{est}$\cite{Kitaev96,Higgins07}. 

Each application of $U$ is controlled by a different qubit which is initially prepared in the state of $\ket{+}=(\ket{0}+\ket{1})/\sqrt{2}$.
The control introduces a phase shift $e^{ip\phi}$ on the $\ket{1}$ component. Measurement takes place in the $\sigma_x$  basis $(X)$ and the results control the additional phase shifts (control phases $\Phi$) on subsequent qubits. This basically enables performing the inverse quantum Fourier transforms without using two-bit or entangled gates \cite{Kitaev96,Higgins07,Griffiths96}. 

As shown by Ref.\cite{Giedke06}, the QPEA does not achieve the SMS that we derived earlier for the Ramsey experiments. However, it solves the problem of needing prior information about the working point. The reason the QPEA does not reach the best sensitivity is due to the fact that for arbitrary phases, we can view the QPEA estimator  $\phi_{est} = 2 \pi (0. u_1 u_2 ... u_K)_2$ as a truncation of an infinite bit string representing the true phase. However, in the QPEA, every control rotation $\Phi_k$ depends on the measurement results of all the bits to the right of the $u_k$ bit. Thus, even if all measurements are perfect, the probabilistic nature of quantum measurements implies that there will be a finite probability to make an error especially for the most significant bits. Although the probability of error is low, the corresponding error is large, and therefore the overall phase variance is increased above the quantum limit. It was noted by Ref.\cite{Giedke06} that by weighting (error-checking) the QPEA for the most significant bits and using fewer measurements on the least significant bits, this problem could be reduced but not completely overcome.

A modified version of the QPEA was introduced by Berry and Wiseman in Ref.\cite{Berry00} which would work for all phases, not just those that were expressible as fractions of powers of two. This model required adaptive control of the phase similar to the QPEA depending on all previous measurements, but also increased the complexity of the calculations. Surprisingly, a simpler version of the Berry and Wiseman algorithm, referred to in Refs.\cite{Higgins09, Said11} as non-adaptive phase estimation algorithm (NAPEA) (\figref{fig:algo}(b)) was also found to give nearly as good results, especially at lower measurement fidelities. In the NAPEA, the number of measurements vary as a function of $k$: $M(K,k)= M_K + F(K-k)$ and the control phase simply cycles through a fixed set of values typically $\Phi = \{ \Phi_1, \Phi_2,\ldots  \Phi_{M(K,k)}\}$ after each measurement. 

Exactly as for the Ramsey measurements, the conditional probability of the $u_m$ measurement is given by: 
\begin{equation}
P(u_m | \phi,k) = \frac{1 + (-1)^{u_m}  D(t, T_2^*) \cos (\phi_k - \Phi)}{2}
\end{equation}
Now, with the assumption of a uniform \emph{a priori} probability distribution for the actual phase $\phi$, Bayes' rule can be used to find the conditional probability for the phase given the next measurement result:
\begin{equation}
P(\phi | \vec{u}_{m+1}) \propto P(u_{m+1} | \phi) P(\phi | \vec{u}_{m})
\end{equation}
where $P(\phi | \vec{u}_{m})\equiv P_m(\phi)  $ is the likelihood function after $u_1, u_2, \ldots u_m$ bit measurements, and gets updated after each measurement. In Refs.\cite{Higgins09, Said11}, the best estimator is again obtained through an integral over this distribution. However, in our work, we simply use the the maximum likelihood estimator ($\phi_{MLE}$) of the  likelihood function\cite{Said11,Higgins09,Berry09}.

In our work, we have chosen to compare the NAPEA with the standard QPEA for several reasons. Firstly, the adaptive algorithm of Refs.\cite{Said11,Higgins09,Berry09} is more difficult to implement experimentally, and in practice seems to offer only slight improvements over the NAPEA. Secondly, the QPEA is a standard PEA which has a simple feed-forward scheme based purely on the bit results. Unless otherwise stated, we set $t_{min}=20$~ns and $T_2^*=1200$~ns in our simulations. Although these values were chosen to be comparable with the typical conditions and limitations in our experimental apparatus, the results and conclusions are valid for any condition in general. The necessary steps involved in the simulation of the both types of PEA are enumerated below.

\subsection{Simulation of PEAs:}
For both QPEA and NAPEA, the following steps are common:
\begin{enumerate} 
\item Parameter initialization: $t_{min}, K, M_K, F, k=K, \Phi=0$
\item Preparation of the initial superposition state: $\rho_0 = \ket{+} \bra{+}$ where $\ket{+}=(\ket{0}+\ket{1})/\sqrt{2}$.
\item The unitary phase operation ${U^2}^{k-1}$ on  $\rho_0$:  $\rho_k ={U^2}^{k-1} \rho_0 {(U^{\dagger})^2}^{k-1}$ \\where $U = \proj{0} +  e^{-i \phi } \proj{1}$ and $\phi=\gamma_e B_{ext} t_{min}$.
\item The feedback rotations $\Phi$ on  $\rho_k$: $\rho_f = R \rho_k R^\dagger$, where $R = \proj{0} +  e^{i \Phi } \proj{1}$.
\item POVM measurement to obtain the signal $S=Tr[M.\rho_f]$ where M is the imperfect measurement operator as described in the text: $M = a \proj{0} +  b \proj{1}$.
\item Assignment of the bit $u_m$ (0 or 1) by comparison of $S$ with the threshold signal.
\end{enumerate}

\noindent \textbf{QPEA:}
\begin{enumerate}
\setcounter{enumi}{6}
\item Repeat steps 2-6 $M(K,k)$ number of times.
\item Update the controls: 
\begin{equation*} \Phi = \sum_{j>k} \frac{u_k}{2^{j-k}} \pi ~, ~~~ k=k-1 
\end{equation*}
where $u_k$ is chosen by majority vote among $\{u_m\}$ for a given $k$.
\item Repeat steps 2-8 until $k=1$. 
\end{enumerate}

\noindent \textbf{NAPEA:}
\begin{enumerate}
\setcounter{enumi}{6}
\item Update the control phase $\Phi$ from the list $\{ \Phi_1, \Phi_2,\ldots  \Phi_{M(K,k)}\}$  
\item Repeat steps 2-7 $M(K,k)$ number of times.
\item Update $k=k-1$.
\item Repeat steps 2-9 until $k=1$.

\end{enumerate}

\section{Results}
\label{sec:results}

\begin{figure}[thb]
	\scalebox{1}[1]{\includegraphics[width=5in]{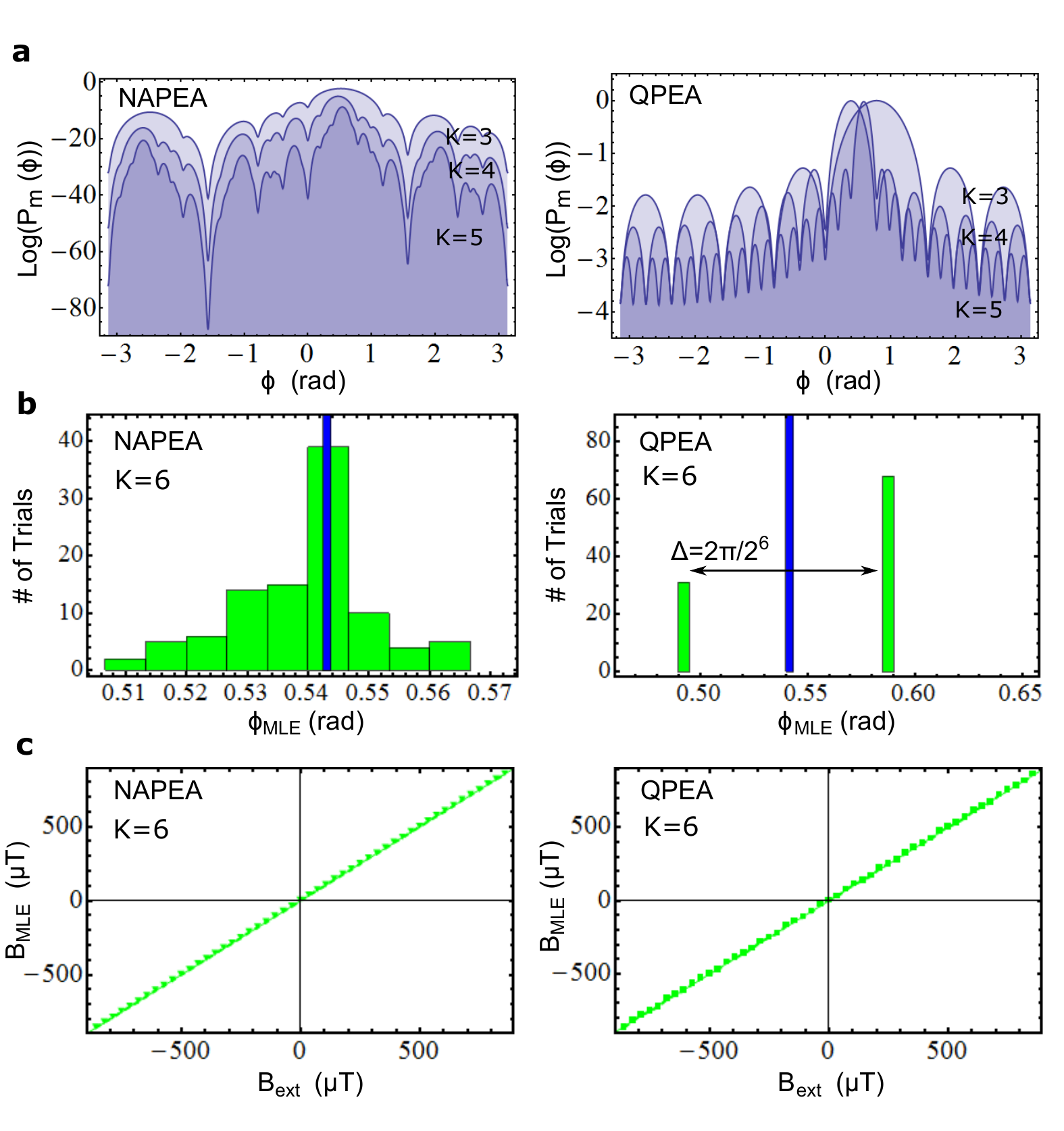}}
	\caption{ \small{ \textbf{Phase Likelihood distribution:} \textbf{a,}  Phase likelihood distribution \textbf{b,} Distribution of MLE's \textbf{c,} Magnetic field readout of QPEA (right) with $\kappa =10\kappa_{th}$, $M_K=1$, $F=0$, $K=6$  and NAPEA (left) $\kappa =5\kappa_{th}$, $M_K=F=4$, $K=6$.}}
	\label{fig:simPEA}
\end{figure}

\figref{fig:simPEA}(a) shows the final phase likelihood distribution as parameter K is varied. The secondary peaks in the final likelihood distribution occur due to the phase ambiguity of individual measurements. As more measurements are performed, these secondary peaks become further suppressed. Note that the figure is given in log scale in order to make the secondary peaks more visible. Recall in QPEA, the bit string itself represents a binary estimate for the unknown phase. However, in order to make a fair comparison between the QPEA and NAPEA, we use the Bayesian approach to analyze the QPEA results as well. The digitization in the phase estimate in QPEA is clearly observed in its phase likelihood distribution. A phase that is perfectly represented by the bit string can lead to a perfect estimate, provided sufficient measurement fidelity is available.

\figref{fig:simPEA}(b) shows the histogram of $\phi_{MLE}$ when each PEA is performed 100 times. While QPEA shows only two possible outcomes for $\phi_{MLE}$, the histogram of $\phi_{MLE}$ for NAPEA is approximately Gaussian around the system phase (blue line). Interestingly, the difference of the two $\phi_{MLE}$ outcomes in QPEA is equal to $2 \pi /2^K $ where $K=6$ in this simulation.

The phase readout $\phi_{MLE}$ is converted to a field readout by the linear relationship: $B_{MLE} =  \phi_{MLE} /  \gamma_e t_{min}$ in  \figref{fig:simPEA}(c) and agrees well with the external magnetic field $B_{ext}$. Hence PEA can be useful for sensing unknown magnetic fields in contrast to the standard Ramsey approach in which the readout is sinusoidally dependent on the external field.

\subsection{Multiple control phases in NAPEA}
\label{sec:multiple}

\begin{figure}[thb]
	\scalebox{1}[1]{\includegraphics[width=5in]{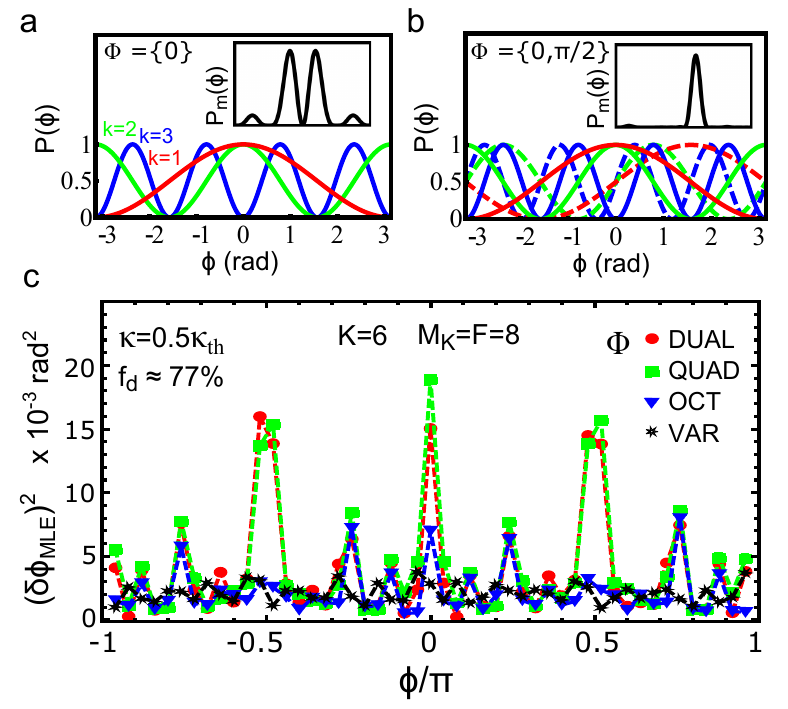}}
	\caption{ \small{ \textbf{Control phases in NAPEA:} \textbf{a,} A simplified form of PEA with a single control phase $\Phi=\{0\}$ leads to a final phase likelihood  distribution which symmetric around ``0'' and consequently gives two possible MLE's.  \textbf{b,} An additional control phase can solve this problem by suppressing the wrong MLE. Red, green and blue lines give the corresponding individual probability distributions for k=1, k=2 and k=3 measurement outcomes respectively. Dashed lines correspond to the case of $\Phi=90^0$ \textbf{c} The red-circle, green-square, blue-triangle, and black-star represent the DUAL, QUAD, OCT, and VAR set of control phases respectively. DUAL and QUAD lead to similar result, and have large variance of the phase readout $(\delta \phi_{MLE})^2$ at points  corresponding to  $\phi \sim 0$ or $ \pm \pi /2$. The OCT suppresses this effect and thus will maintain almost a constant sensitivity over the full field range. The VAR leads to further improvement in consistency because of the rapid increment in the number of control phases according to the weighting scheme. The standard deviation of the phase readout variances at all points $\sigma_{(\delta \phi_{MLE})^2}$, corresponding to DUAL, QUAD, OCT and VAR sets are  4.13, 4.28, 1.80 and 0.71,  10$^{-3}$rad$^2$  respectively.}}
	\label{fig:compDQOV}
\end{figure}

To understand the choice of control phases in the NAPEA, one could imagine a simple version of NAPEA without multiple control phases, i.e., $\Phi=\{0\}$. The final phase likelihood  distribution in this case will be symmetric about the origin $\phi = 0$ (see \figref{fig:compDQOV}(a) inset), because the likelihood distribution is a product of many even cosine functions. Introducing a second control phase  i.e., $\Phi=\{0, \pi/2 \}$ breaks this symmetry and result in a unique answer for $\phi_{MLE}$(\figref{fig:compDQOV}(b)). However, introducing even more control phases can be useful for obtaining a consistent sensitivity throughout the full field range. Table~\ref{controlphases} summarizes the terms that will be used in this paper, in describing the different sets of control phases. 

\begin{table}[h!]
\centering
    \begin{tabular}{ | l | l | l |}
    \hline
    Control phases $\{\Phi\}$    & Term & $\sigma_{(\delta \phi_{MLE})^2}$ (rad$^2$) \\ \hline
    $\{0, \frac{\pi}{2} \}$               & DUAL & 4.13$\times 10^{-3}$  \\ \hline
    $\{0, \frac{\pi}{2}, \pi, \frac{3 \pi}{2} \}$ & QUAD & 4.28$\times 10^{-3}$  \\ \hline
    $\{0, \frac{\pi}{4}, \frac{\pi}{2}, \frac{3\pi}{4}, \pi, \frac{5\pi}{4}, \frac{3\pi}{2} , \frac{7\pi}{4} \}$ & OCT & 1.80$\times 10^{-3}$ \\ \hline   
    $\{0, \frac{\pi}{{M(K,k)}}, \frac{2\pi}{{M(K,k)}}, \ldots \frac{[{M(K,k)}-1]\pi}{{M(K,k)}} \}$ & VAR & 0.71$\times 10^{-3}$ \\ \hline 
    \end{tabular}
\caption{List of control phases used in NAPEA and corresponding terms used in the paper. }\label{controlphases}
\end{table}

The variance of the phase readout $(\delta \phi_{MLE})^2$ with respect to the given quantum phase is plotted in \figref{fig:compDQOV}(c). It is noteworthy that a QUAD set of control phases is no better than the DUAL set. While former case leads to X and Y basis measurements, latter case corresponds to X, Y, -X and -Y basis measurements. Therefore similarity in results of DUAL and QUAD sets could be explained as follows. Imagine a condition that resulted in a bit measurement $u_m$ in the X(Y) basis. The same condition would have resulted a bit measurement $1-u_m$ in the -X(-Y) basis which will eventually result in the same probability distributions. Because the DUAL and the QUAD sets implies measurement in $\{X,Y\}$ and $\{X,Y,-X,-Y\}$ basis respectively, they tend to the same final results, and are technically equivalent. As seen in \figref{fig:compDQOV}(c), the DUAL and the QUAD cases have relatively worse phase variance at working points corresponding to $\phi \sim 0$ or $ \pm \pi /2$. This effect is significantly suppressed in the case of eight control phases, the OCT set. Using the variable set of control phases (VAR) leads to further improvement in consistency because of the rapid increment in the number of control phases according to the weighting scheme. However, the VAR set can be comparatively difficult to implement in practice. The consistency of the various sets  of control phases are summarized in Table~\ref{controlphases} by calculating the standard deviation of the variance over the entire interval $(-\pi,\pi)$.

\subsection{Weighting scheme and the measurement fidelity}

\begin{figure}[thb]
	\scalebox{1}[1]{\includegraphics[width=6in]{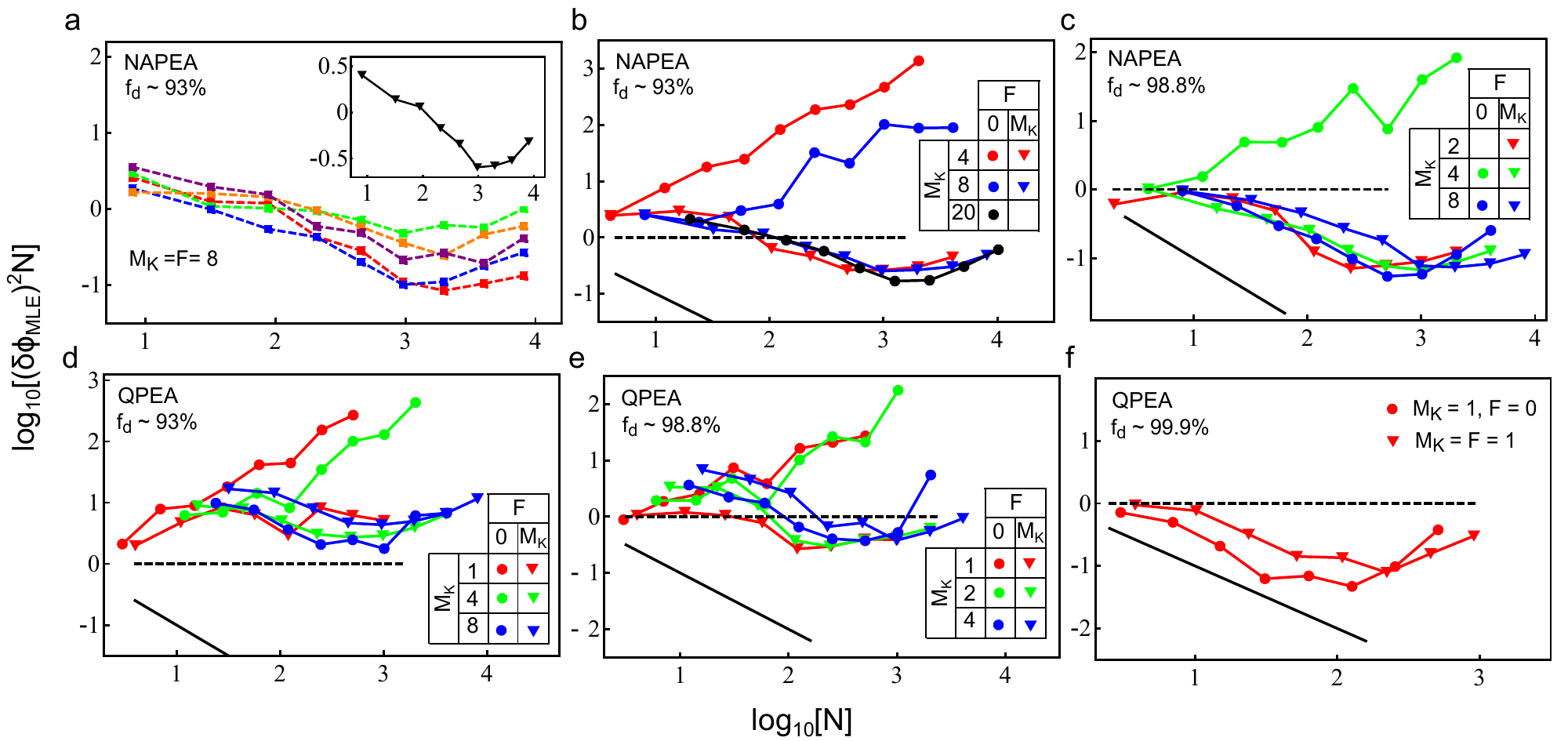}}
	\caption{ \small{ \textbf{Precision scaling of the phase measurement} \textbf{a,} The NAPEA sensitivity $(\delta \phi_{MLE})^2 N$ obtained as resource is increased by increment of $K$ for different system phases. (inset) The average obtained for above system phases is shown. \textbf{b-f,} The precision scaling of average NAPEA (QPEA) phase sensitivity under different conditions of fidelity and PEA parameters are shown in \textbf{b-c,} (\textbf{d-f}). $F = 0$ implies no weighting. The best results in QPEA requires extremely high fidelity $> 99\%$ whereas for NAPEA a fidelity $\approx 90\%$ is sufficient. The black dashed (solid) line shows standard-quantum (Heisenberg) limit for phase measurement.}}
	\label{fig:peascaling}
\end{figure}

In this section, we explore the effect of weighting scheme and the measurement fidelity on the NAPEA and QPEA. \figref{fig:peascaling} gives phase sensitivity scaling $(\delta \phi_{MLE})^2 N$ when $K$ is increased from 1 to 9. Here $N$ is equivalent to the number of unitary operations in Refs.\cite{Higgins09, Said11} and can be calculated as below.
\begin{equation}
\label{eq:T}
N = \sum_{k=1}^K  M(K,k) \, 2^{(k-1)}  \, = [M_K (2^K -1) + F (2^K -K -1)]
\end{equation}
\figref{fig:peascaling}(a) shows the scaling of sensitivity with $N$ for five different choices of quantum phases $\phi= (\pi/9.789, \pi/7.789, \pi/5.789, \pi/3.789, \pi/1.789) $, while the inset figure gives the average behavior. Here onward, we present only the average result in the scaling plots for clarity. It is important to present the average behavior rather than the behavior for a particular quantum phase, because the sensitivity is not necessarily the same for all phases as shown in the previous section. From \figref{fig:peascaling}(b) and (c), it is clear that although weighting can play a role in NAPEA, there also exist non-weighted choice of optimal results. However, the optimal non-weighted parameters are highly dependent on $f_d$. For instance, with $f_d = 98.8\%$, the optimum non-weighted parameters were found to be \{$K, M_K, F$\}=\{7,8,0\} while the same parameters led to  $\sim 10^3$ worse sensitivity when $f_d = 92.9\%$. On the other hand the weighted parameters \{7,8,8\}  resulted in nearly same sensitivities in either case.

In QPEA, the change in control phase $\Phi$ occurs with the change in $k$. Moreover, only a single bit measurement result $u_k$ is available for each $k$, unlike in the case of NAPEA where there exist $M(k,K)$ bit measurements. However, in order to make a fair comparison, we still perform the weighting scheme on QPEA as described in section \ref{sec:pea} to obtain $M(k,K)$ bit measurements. We use majority voting of bit measurements for determination of the control phases. It turns out that the best results in QPEA are obtained only with extremely high measurement fidelity ($f_d > 99 \%$) and requires no weighting ($M_K=1, F=0$). Further, even after using Bayesian estimation, the sensitivity in QPEA is ultimately limited by the minimum bit error of the phase readout given by $\delta \phi_{est} = \pi/2^K$. 

\subsection{Field sensitivity and PEA performance}
\label{sec:cmpsensdyrg}

\begin{figure}[thb]
	\scalebox{1}[1]{\includegraphics[width=6.5in]{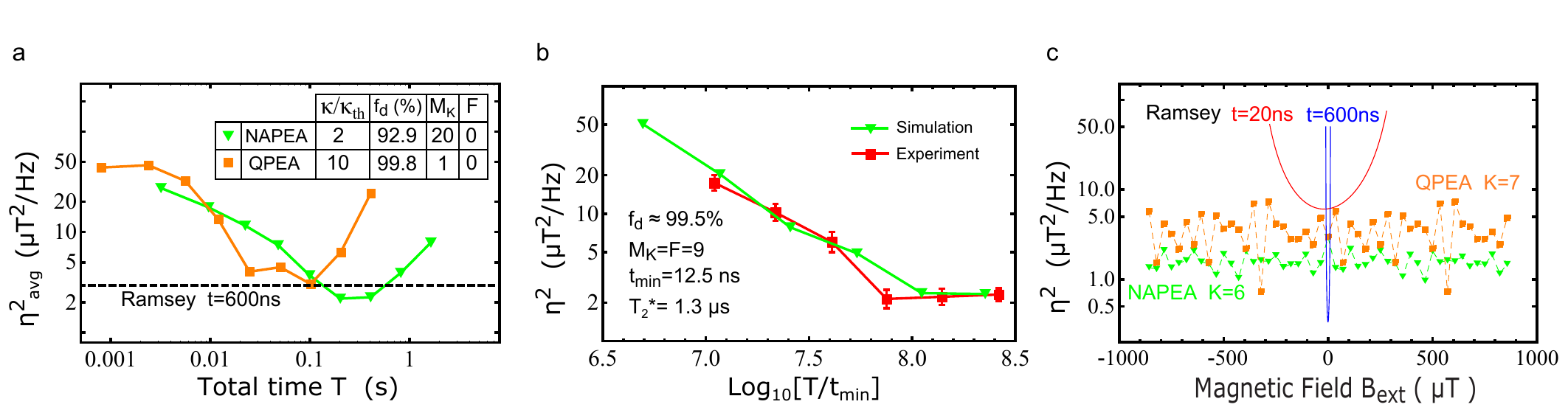}}
	\caption{\small{ \textbf{Magnetic field sensitivity} \textbf{a,} Field sensitivity $\eta^2 = (\delta B)^2 T$ averaged over many different field with the number of resources. The resources is increased by the increment of K and given in units of time. The best cases obtained from QPEA and NAPEA is compared with the optimum Ramsey sensitivity averaged over its full field range. \textbf{b,} Field sensitivity with unitless number of resources $N=T/t_{min}$. The simulation here was performed with the exact experimental conditions and is in well agreement with experimental results. \textbf{c,} The NAPEA sensitivity is roughly constant throughout the full field range whereas QPEA sensitivity is not. The red (blue) curves show theoretical limits of Ramsey sensitivity for t=20~ns (t=600~ns).}}
	\label{fig:fieldsensitivity}
\end{figure}

The corresponding scaling of field sensitivity $\eta^2 = (\delta B)^2 T $ for some of the data in \figref{fig:peascaling} is shown in \figref{fig:fieldsensitivity}(a). Here, $(\delta B)^2 = \langle (B_{MLE}-B_{ext})^2 \rangle$ is the variance of field with respect to the external magnetic field and $T = N t_{min} \kappa$ is the total evolution time of the PEA. The best results from NAPEA was obtained with a fidelity 92.9\% and an OCT set of control phases.  QPEA's best results requires extremely high fidelity 99.9\%, and furthermore show a significant fluctuation in the sensitivity over the full field range. The statistics obtained here along with PEA parameters used are summarized in the Table~\ref{bestresults}.
\begin{table}
\centering
    \begin{tabular}{ | l | l | l | l | l | l |}
    \hline
    PEA & Fidelity &Parameters &$\sigma_{(\delta \phi_{MLE})^2}$ (rad$^2$)&  $\eta^2_{avg}$ ($\mu T^2/Hz$) &  $T$ (s)\\ \hline
    NAPEA(OCT) & 92.9\% ($\kappa =2\kappa_{th}$) & $M_K=20, F=0, K=6$ & 0.022$\times 10^{-3}$ & $1.58 \pm 0.35$ & 0.202  \\ \hline
    QPEA & 99.9\%($\kappa =10\kappa_{th}$)& $M_K = 1, F=0, K=7$ & 0.197$\times 10^{-3}$ & $3.67 \pm 1.62 $ & 0.102  \\ \hline
    \end{tabular}
    \caption{Summery of best results from QPEA and NAPEA }\label{bestresults}
\end{table}

While NAPEA demands relatively lesser measurement fidelity than in QPEA, the total estimation time is larger. However, as shown in above table and \figref{fig:fieldsensitivity}(c), the sensitivity obtained from NAPEA is better and more consistent compared to QPEA. Although it is possible to enhance the dynamic range of PEA by reducing $t_{min}$ and thereby achieving higher $K$,  no significant improvement in sensitivity was observed because it is ultimately limited by the SMS at the longest evolution time. On the other hand reaching the best SMS in Ramsey limits the dynamic range. 

\begin{figure}[thb]
	\scalebox{1}[1]{\includegraphics[width=5in]{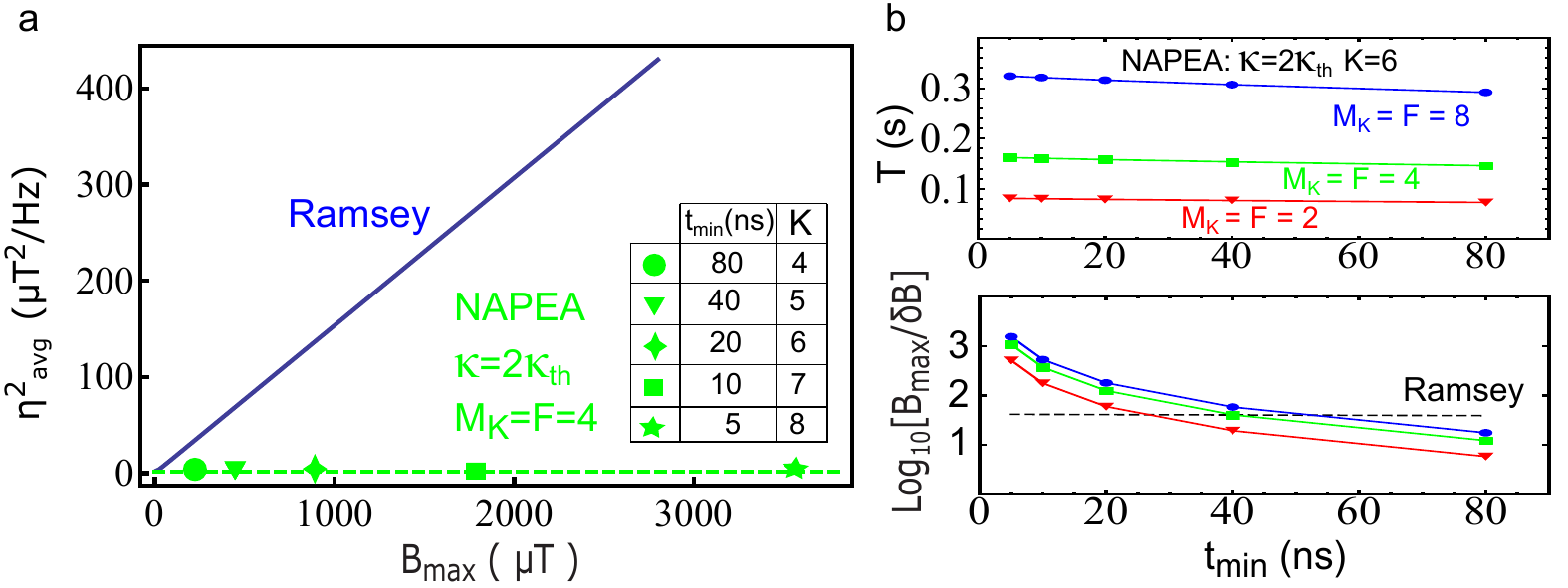}}
	\caption{\small{ \textbf{Dynamic range and Time Constants} \textbf{a,} The standard Ramsey approach show a trade-off between the sensitivity and $B_{max}$ whereas PEA can maintain the same sensitivity upto a wide field range. \textbf{b,} (top) The total time for the NAPEA as a function of minimum evolution time interval $t_{min}$. The parameter K is chosen such that the longest evolution time interval is always the same, 1280~ns. (bottom) The corresponding dynamic range $DR=B_{max}/\delta B$ is shown, where $\delta B$ is the minimum detectable field change. The black dashed line is obtained from the Ramsey theoretical limit $\eta_{th} / \sqrt{T}$ for the same experimental conditions. Here, the total time for PEA with $M_K=F=8$ is used for T.}}
	\label{fig:drtc}
\end{figure}

\figref{fig:drtc}(b) show the total time and the dynamic range $DR=B_{max}/\delta B$ as a function of $t_{min}$ for different choices of NAPEA parameters: $M_K$ and $F$. The parameter K is chosen such that the longest evolution time interval is always the same i.e., $2^{K-1} t_{min} \approx T_2^*$. Here, $\delta B$ is the minimum detectable field amplitude. The corresponding Ramsey DR obtained for an averaging time equal to that of NAPEA with $M_K=F=8$ is also shown. Clearly, NAPEA gives better DR for smaller $t_{min}$. By a suitable choice of NAPEA parameters we can reduce the time constant without significant compromise between the sensitivity and DR. For instance, when $t_{min}$=10~ns, a change in the NAPEA parameters from $M_K=F=8$ to $M_K=F=4$ will reduce the time constant by 50\% though the reduction in sensitivity and DR is only $\sim 28\%$. 

In principle, $t_{min}$ could be lowered to any value in order to achieve a desired dynamic range. However in practice, this is limited by the finite pulse length and gives a lower-bound; $t_{min} > t_{\pi}$. On the other hand, strong qubit driving can invalidate the RWA due to the effect known as the Bloch-Siegert shift\cite{Tuorila10, Fuchs09}. Here, the qubit resonance is shifted by a factor of $(1+\Omega^2/4 \omega_0^2)$ in the rotating frame of the driving field where $\Omega$ is the Rabi frequency and $\omega_0$ is the qubit resonance frequency in the Lab frame. However, the RWA can still be reasonably applicable upto a $\sim 1\%$ of a Bloch-Siegert shift\cite{Fuchs09} corresponding to $\Omega / \omega_0 \sim 1/5$. This suggests that, in our application where a background magnetic field of $\sim$~500~G leads to a qubit frequency $\omega_0 \approx 1.4$~GHz associated with $m_s=0 \leftrightarrow m_s=-1$ transition, the Rabi driving could be made as strong as $\Omega \sim 300$~MHz resulting to lower bound of $t_{min} \sim 3.4$~ns. In case of driving the $m_s=0 \leftrightarrow m_s=+1$ transition under the same conditions, qubit frequency is $\omega_0 \approx 4.3$~GHz and corresponds to a lower bound of $t_{min} \sim 1.2$~ns. Extrapolation from \figref{fig:drtc}(b) data gives the upper bound for the dynamic range in this case, $DR \sim 10^5$, which should be sufficient to simultaneously detect the fields from both electron and nuclear spins in a single magnetic field image.

\subsection{PEA for AC Magnetometry}
\label{sec:acpea}

The best sensitivity in DC magnetometry is limited by the dephasing time $T_2^*$ which is usually much less than the decoherence time $T_2$. Therefore, one could be interested in implementing the PEA for AC magnetometry in order to achieve improvement in the sensitivity: $\eta_{AC}\approx \eta_{DC} \sqrt{\frac{T_2^*}{T_2}}$. Here we show by simulations, how PEA could be applied for sensing AC magnetic fields, $b(t)=b_{ac} \cos(\omega t + \theta)$. Our approach can be used to sense an unknown field amplitude $b_{ac}$ as well as the phase $\theta$ of the field. Because our focus is only to describe the method of implementation, we consider the ideal scenario of 100\% photon efficiency and neglect the effect of decoherence for simplicity.

\begin{figure}[thb]
	\scalebox{1}[1]{\includegraphics[width=5in]{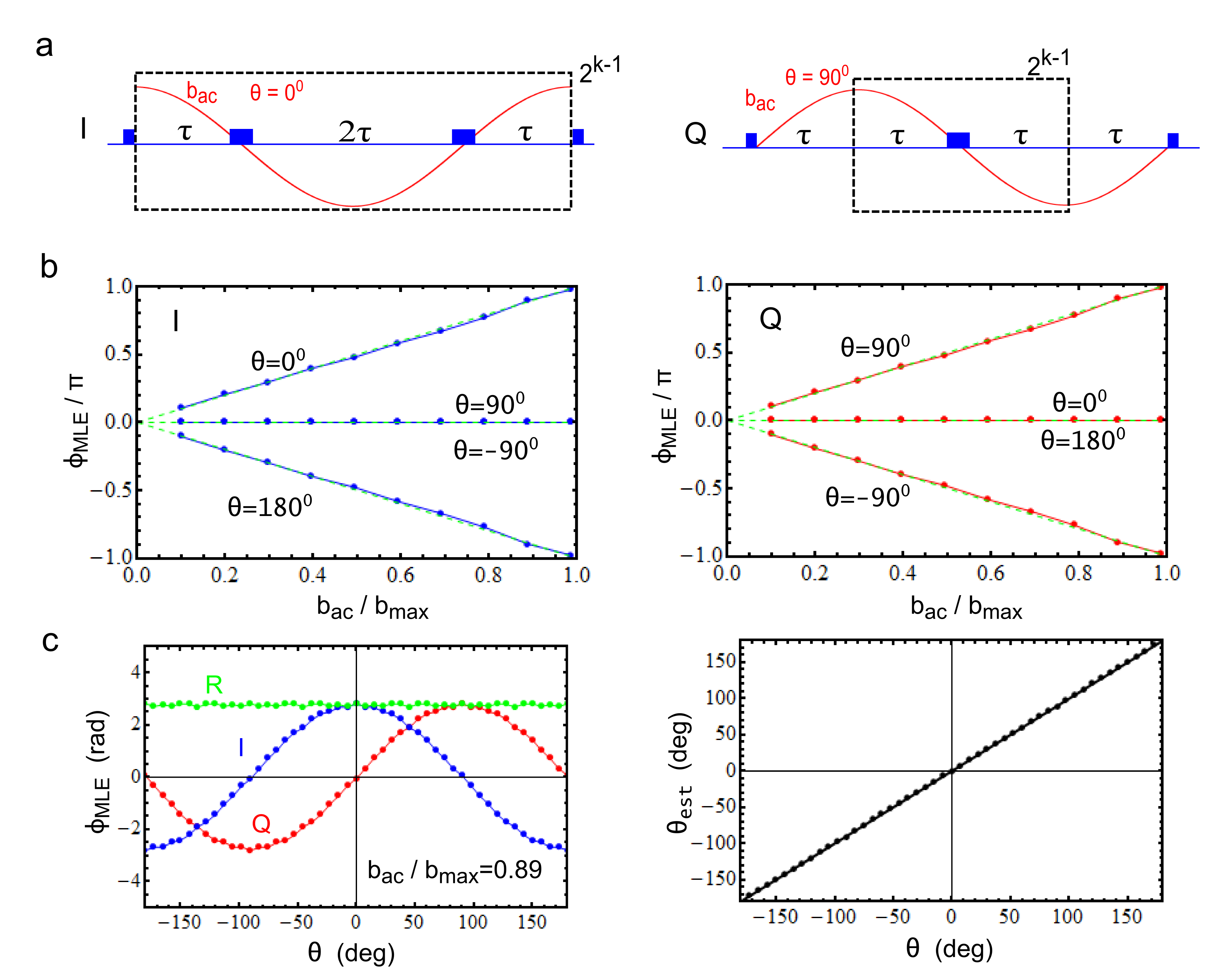}}
	\caption{\small{ \textbf{AC magnetometry with PEA}  \textbf{a,} type-I (left) and type-Q (right) MW pulse sequence used for necessary phase accumulations from the AC magnetic field. \textbf{b,} PEA readout with type-I (left) and type-Q (right) sequences. \textbf{c,} detection of the AC magnetic field phase.}}
	\label{fig:acfig}
\end{figure}

Performing PEA requires the ability to accumulate several phases: $\phi, 2\phi, 4\phi$ etc, where $\phi$ is the unknown quantum phase to be measured. In DC magnetometry, these phase accumulations are achieved by varying the free evolution time in Ramsey sequence. In order to achieve the required phase accumulations from an AC-field, we can have two types of echo-based pulse sequences referred to as type-I and type-Q (\figref{fig:acfig}(a)) in this paper. Type-I sequence is maximally sensitive to magnetic fields with $\theta=0^0$ or $\theta=180^0$ whereas completely insensitive (i.e., gives zero phase accumulation) to $\theta=\pm 90^0$. Type-Q sequence on the other hand, is maximally sensitive to magnetic fields with $\theta=\pm 90^0$  and completely insensitive to  $\theta=0^0$ or $\theta=180^0$ (\figref{fig:acfig}(b)). Further, a magnetic field with an arbitrary phase $\theta$ could be expanded as: 
\[b(t)=b_{ac} \cos(\omega t + \theta) = b_{ac} \cos(\theta) \cos(\omega t)-b_{ac} \sin(\theta) \sin(\omega t) \]
Therefore PEA with type-I and  type-Q sequences lead to readout $\phi_I \propto b_{ac} \cos(\theta) $ and $\phi_Q \propto b_{ac} \sin (\theta)$ respectively. Hence, the phase information of the unknown field could be extracted: $\theta_{est} = \tan^{-1}[ \phi_Q / \phi_I ]$  (\figref{fig:acfig}(c)). Application of PEA for AC magnetometry has recently been demonstrated. For further details see Ref.\cite{Nusran13} and its online supplementary material.


\section{Conclusion}
\label{sec:conclusion}

In conclusion we have made a detailed investigation of PEA approach for magnetic field sensing via Monte-Carlo simulations and compared with Ramsey magnetometry. The importance of dynamic range for magnetic imaging of unknown samples was also emphasized. The high dynamic range and the linear response to the field amplitude makes PEA useful for many practical applications. When it comes to NAPEA, DUAL and QUAD set of control phases give similar results and have relatively worse sensitivities at working points corresponding to $\phi \sim 0$ or $ \pm \pi /2$. This effect can be suppressed by introducing more control phases. In particular, the use of OCT case set of control phase lead to a significant improvement in uniformity of the sensitivity over the full field range. The weighting scheme can play a role in NAPEA but not in QPEA. Even for NAPEA, there is always a choice of non-weighted PEA parameters that can lead to optimum results, but the optimum parameters in general depend on the measurement fidelity. The best results in NAPEA are however guaranteed for measurement fidelity above $\sim 90\%$. QPEA shows a significant variation in the sensitivity across the full field range as a consequence of the binary error in the readout. Further, the best results in QPEA demands extremely high fidelity $\sim 99\%$. Because multiple measurements are not required, the total estimation time for QPEA is much less than in NAPEA. In any case, NAPEA seems to be superior to QPEA due to (a) better sensitivity, (b) consistency in sensitivity throughout the full field range, (c) comparatively less demanding measurement fidelity and (d) for its simplicity in experimental realization. Finally, we have shown that PEA can also be implemented for detection of unknown AC magnetic fields. Our method allows for the detection of both field amplitudes, and the phase of the field.

\section{Acknowledgements}

This work was supported by the Department of Energy Award No. DE-SC0006638 for optimization of magnetometry techniques, key equipment, materials and effort by N.M.N and G.D.; NSF Award No. DMR-0847195 for development of experimental setup, NSF Award No. PHY-1005341 for development of fabrication techniques and collaborations. G.D. gratefully acknowledges support from the Alfred P. Sloan Foundation.


\newpage

%

\bibliographystyle{apsrev4-1}

\newpage

\newpage

\newpage

\newpage

\newpage

\newpage

\newpage

\newpage

\newpage

\newpage

\end{document}